\documentclass[twocolumn,showpacs,preprintnumbers,amsmath,amssymb,superscriptaddress]{revtex4}

\usepackage{graphicx}
\usepackage{dcolumn}
\usepackage{bm}

\begin{document}
\draft
\title{Synthesis of superheavies: State of affairs and outlooks}

\author{Valery~Zagrebaev}
\affiliation{Flerov Laboratory of Nuclear Reactions, JINR, Dubna,
Moscow Region, Russia}
\author{Walter~Greiner}
\affiliation{Frankfurt Institute for Advanced Studies, J.W.
Goethe-Universit\"{a}t, Frankfurt, Germany}

\date{\today}

\begin{abstract}
Nuclear reactions leading to formation of new superheavy elements
and isotopes are discussed in the paper. ``Cold'' and ``hot''
synthesis, fusion of fission fragments, transfer reactions and
reactions with radioactive ion beams are analyzed along with their
abilities and limitations. Several most promising reactions are
proposed for experimental study.
\end{abstract}
 \maketitle

\section{Introduction}

Two important pages in synthesis of superheavy (SH) nuclei have
been overturned within last twenty years. In the ``cold'' fusion
reactions based on the closed shell target nuclei, lead and
bismuth, SH elements up to $Z=113$ have been produced
\cite{Hoffman00,Morita}. The ``world record'' of 0.03~pb in
production cross section of 113 element has been obtained here
within more than half-year irradiation of $^{209}$Bi target with
$^{70}$Zn beam \cite{Morita}. Further advance in this direction
(with Ga or Ge beams) seems to be very difficult. Note also that
the SH elements obtained in the ``cold'' fusion reactions with Pb
or Bi target are situated along the proton drip line being very
neutron-deficient with a short half-life.

The cross sections for SH element production in more asymmetric
(and ``hoter'') fusion reactions of $^{48}$Ca with actinide
targets were found much larger \cite{Ogan07}. Even 118 element was
produced with the cross section of about 1~pb in the
$^{48}$Ca+$^{249}$Cf fusion reaction \cite{Ogan06}. Fusion of
actinides with $^{48}$Ca leads to more neutron-rich SH nuclei with
much longer half-lives. However they are still far from the center
of the predicted ``island of stability'' formed by the neutron
shell around $N=184$ (these are the $^{48}$Ca induced fusion
reactions which confirm an existence of this ``island of
stability''). Moreover, californium is the heaviest actinide which
can be used as a target material in this method (the half-life of
the most long-living einsteinium isotope, $^{252}_{99}$Es, is 470
days, sufficient to be used as target material, but it is
impossible to accumulate required amount of this matter).

In this connection other ways for the production of SH elements
with $Z>118$ and also neutron rich isotopes of SH nuclei in the
region of the ``island of stability'' should be searched for. In
this paper we analyze abilities and limitations of different
nuclear reactions leading to formation of SH elements (``cold''
and ``hot'' synthesis, symmetric fusion, transfer reactions and
reactions with radioactive beams) trying to find most promising
reactions which may be used at available facilities.

\section{The model}\label{Model}

The cross section of SH element production in heavy ion fusion
reaction (with subsequent evaporation of $x$ neutrons in the
cooling process) is calculated as follows

\begin{equation}
\sigma _{\rm ER}^{\rm xn}(E)=\frac{\pi}{k^2} \sum\limits_{l =
0}^\infty (2l + 1)P_{\rm cont}(E,l)\cdot P_{\rm CN}(E^*,l) \cdot
P_{\rm xn} (E^*,l). \label{Sig}
\end{equation}

Empirical or quantum channel coupling models \cite{NRV} may be
used to calculate rather accurately penetrability of the
multi-dimensional Coulomb barrier $P_{\rm cont}(E,l)$ and the
corresponding capture (sticking) cross section, $\sigma _{\rm
cap}(E)=\pi/k^2 \sum (2l+1) P_{\rm cont}$. The survival
probability $P_{\rm xn}(E^*)$ of an excited compound nucleus (CN)
can be calculated within a statistical model. We use here the
fission barriers and other properties of SH nuclei predicted by
the macro-microscopic model \cite{Moller95}. Other parameters
determining the decay widths and the algorithm itself for a
calculation of the light particle evaporation cascade and $\gamma$
emission are taken from \cite{Zag02}. All the decay widths may be
easily calculated also at the Web site \cite{NRV}.

The probability for compound nucleus formation $P_{\rm CN}(E,l)$
is the most difficult part of the calculation. In \cite{Zag01} the
two-dimensional master equation was used for estimation of this
quantity, and a strong energy dependence of $P_{\rm CN}$ was
found, which was confirmed recently in experiment \cite{Naik07}.
Later the multi-dimensional Langevin-type dynamical equations were
proposed \cite{ZG05,ZG07} for the calculation of the probability
for CN formation both in ``cold'' and ``hot'' fusion reactions.
The main idea is to study evolution of the heavy nuclear system
driven by the time dependent multi-dimensional potential energy
surface gradually transformed to the adiabatic potential
calculated within the two-center shell-model \cite{TCSM}. Note
that the extended version of this model developed recently in
\cite{extTCSM} leads to a correct asymptotic value of the
potential energy of two separated nuclei and height of the Coulomb
barrier in the entrance channel (fusion), and appropriate behavior
in the exit channel, giving the required mass and energy
distributions of reaction products and fission fragments.

\begin{figure}[t]
\begin{center}
\includegraphics[width = 6.0 cm, bb=0 0 1043 2539]{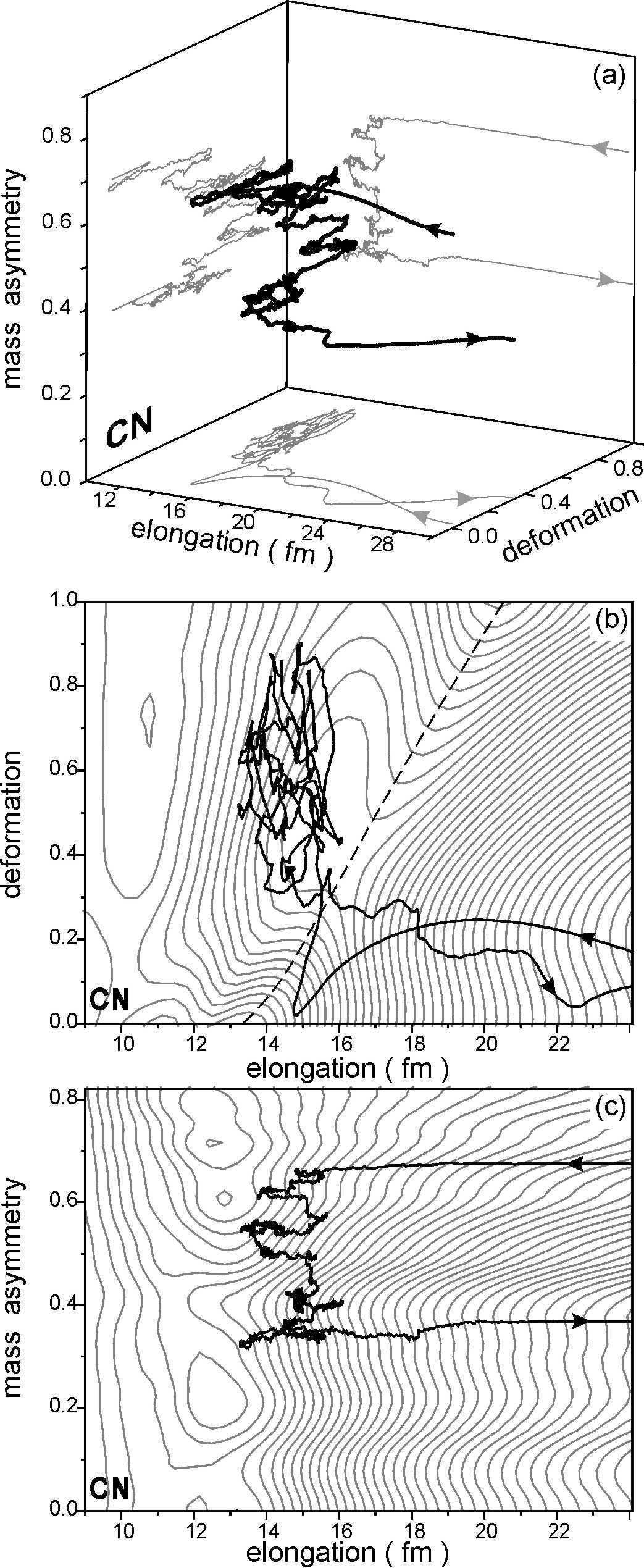} \end{center}
\caption{Collision of $^{48}$Ca +$ ^{248}$Cm at
$E_{c.m.}=210$~MeV. One of typical trajectories calculated within
the Langevin equations and going to the quasi-fission exit channel
(lead valley) is shown in the three-dimensional space (a) and
projected onto the ``deformation--elongation'' (b) and
``mass-asymmetry--elongation'' (c) planes. The dashed line in (b)
shows the ridge of the multidimensional Coulomb
barrier.\label{traj}}
\end{figure}

In the case of near-barrier collision of heavy nuclei only a few
trajectories (of many thousands tested) reach the CN configuration
(small values of elongation and deformation parameters, see
Fig.~\ref{traj}). All others go out to the dominating deep
inelastic and/or quasi-fission exit channels. One of such
trajectories is shown in Fig.~\ref{traj} in the three-dimensional
space of ``elongation--deformation--mass-asymmetry'' used in the
calculations.

Made within our approach the predictions for the excitation
functions of SH element production with Z=112$\div$118 in
1n$\div$5n evaporation channels of the $^{48}$Ca induced fusion
reactions \cite{Zag04a,Zag04b} agree well with the later obtained
experimental data. This gives us confidence in receiving rather
reliable estimations of the reaction cross sections discussed
below. Such estimations are urgently needed for planning future
experiments in this field.

\section{Cold fusion reactions}\label{Cold}

\begin{figure}[t]
\begin{center}
\includegraphics[width = 8.0 cm, bb=0 0 1382 778]{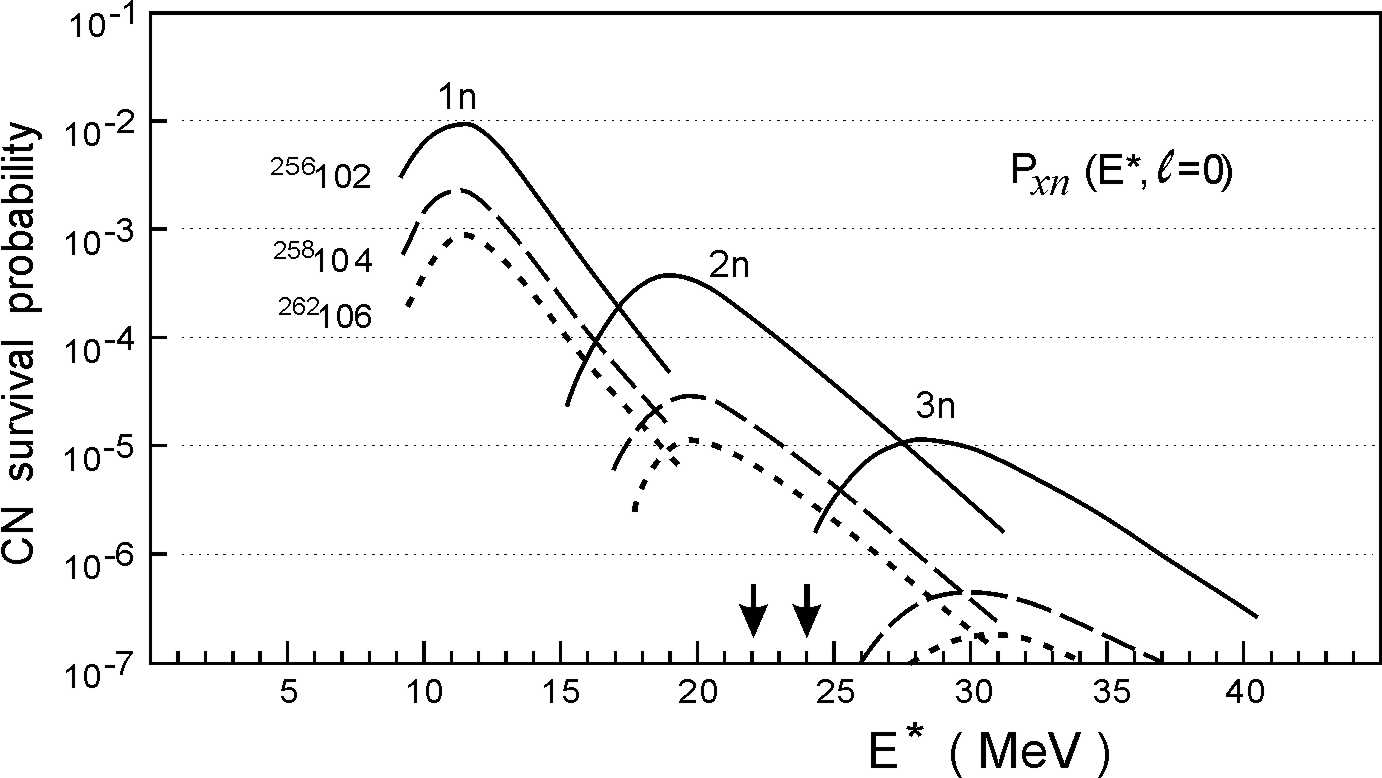} \end{center}
\caption{Survival probability $P_{\rm xn}(E^*,l=0)$ of $^{256}$No,
$^{258}$Rf and $^{262}$Sg compound nuclei produced in the
$^{48}$Ca+$^{208}$Pb, $^{50}$Ti+$^{208}$Pb and
$^{54}$Cr+$^{208}$Pb fusion reactions. The arrows indicate the
Bass barriers (see Table~\ref{bf}).\label{pxn}}
\end{figure}

At near-barrier incident energies fusion of heavy nuclei
($^{48}$Ca, $^{50}$Ti, $^{54}$Cr and so on) with $^{208}$Pb or
$^{209}$Bi targets leads to formation of low-excited superheavy CN
(``cold'' synthesis). In spite of this favorable fact (only one or
two neutrons are to be evaporated), the yield of evaporation
residues sharply decreases with increasing charge of synthesized
SH nucleus. There are two reasons for that. First, in these
reactions neutron deficient SH nuclei are produced far from the
closed shells or sub-shells. As a result, neutron separation
energies of these nuclei are rather high whereas the fission
barriers (macroscopic components plus shell corrections) are
rather low (see Table~\ref{bf}). This leads to low survival
probability even for 1n and 2n evaporation channels,
Fig.~\ref{pxn}.

\begin{table}[h]
\caption{Fission barriers (macroscopical part and shell
correction) and neutron separation energies (MeV) of CN produced
in the $^{48}$Ca+$^{208}$Pb, $^{50}$Ti+$^{208}$Pb and
$^{54}$Cr+$^{208}$Pb fusion reactions \cite{Moller95}. The last
column shows the excitations of CN at the Bass barrier
\cite{Bass80} incident energies. \label{bf}}
\begin{tabular}{|c|c|c|c|c|c|}
\hline
CN & B$_{\rm LD}$ & Sh.Corr. &  $B_{\rm fis}$ & $E_n^{\rm sep}$ & $E^*$(Bass)\\[2pt]
\hline
$^{256}$No & 1.26 & 4.48 & 5.7  & 7.1 & 22  \\[2pt]
$^{258}$Rf & 0.71 & 4.49 & 5.3  & 7.6 & 24  \\[2pt]
$^{262}$Sg & 0.47 & 4.63 & 5.1  & 7.8 & 24  \\[2pt]
\hline
\end{tabular}
\end{table}

\begin{figure}[t]
\begin{center}
\includegraphics[width = 8.0 cm, bb=0 0 1253 2153]{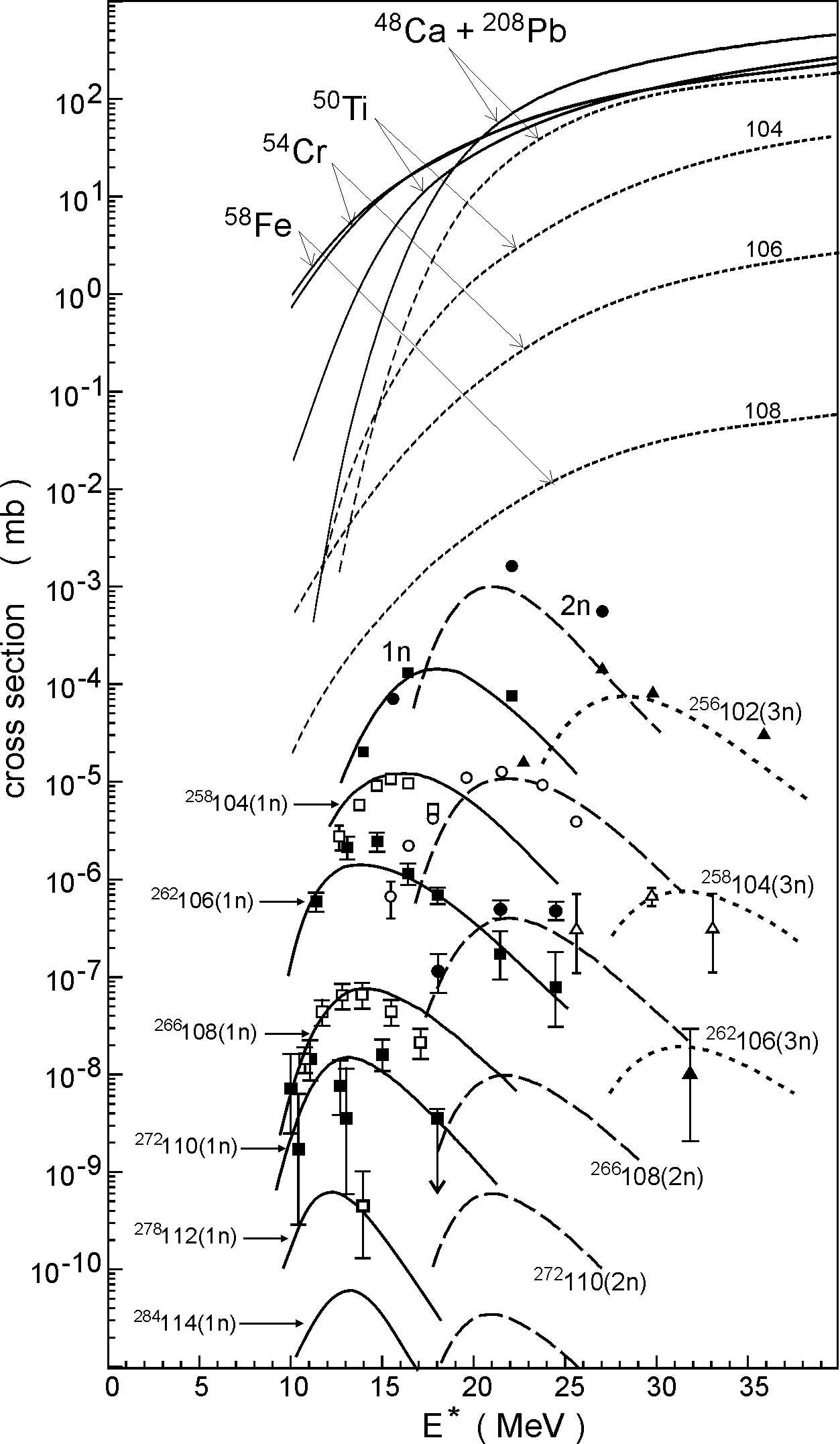} \end{center}
\caption{Capture (upper solid curves), CN formation (short-dashed
curves) and SH element production cross sections in the $^{208}$Pb
induced fusion reactions. 1n, 2n and 3n evaporation channels are
shown by solid, dashed and dotted curves (theory) and by
rectangles, circles and triangles (experiment), correspondingly.
Experimental data are taken from \cite{Hoffman00,Morita,Yer98}.
\label{cold_cs}}
\end{figure}

\begin{figure}[t]
\begin{center}
\includegraphics[width = 8.0 cm, bb=0 0 1254 883]{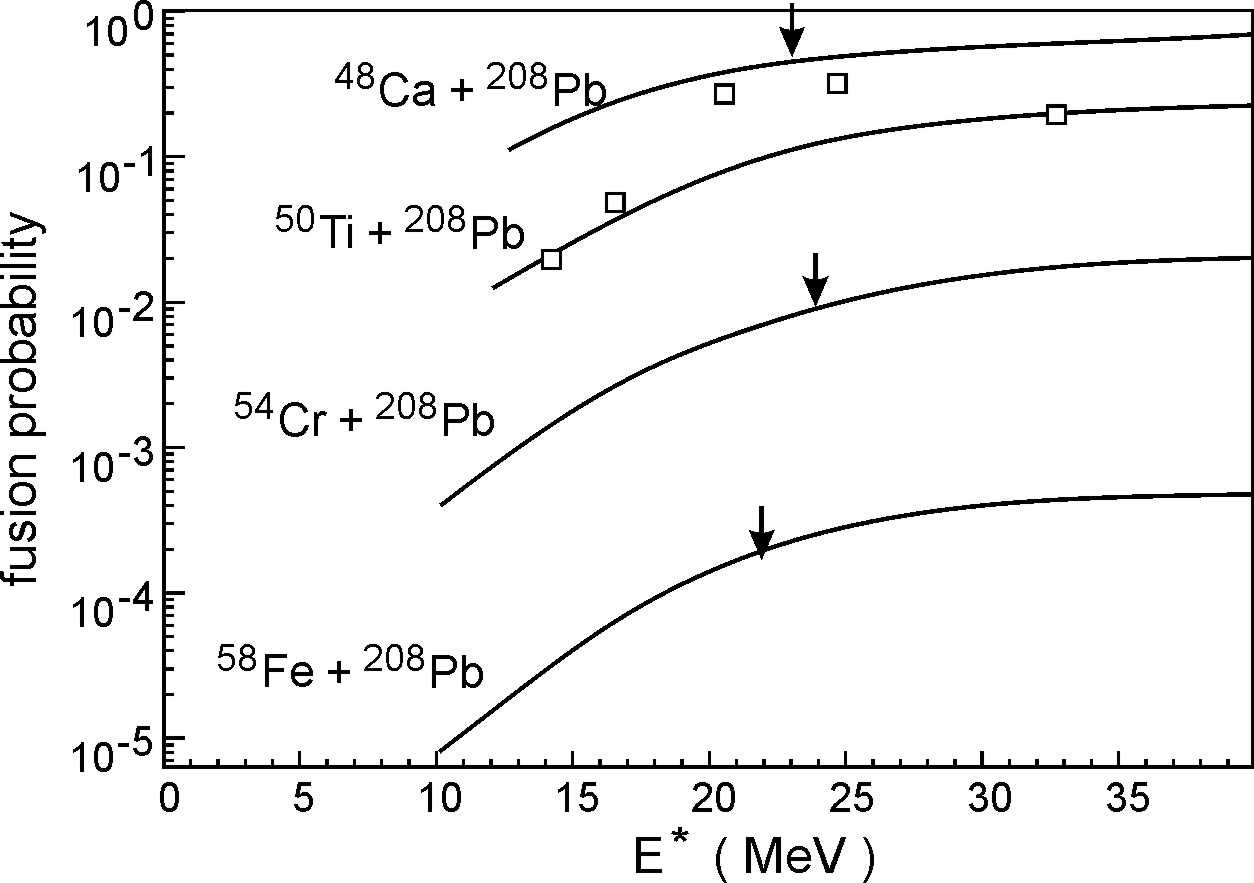} \end{center}
\caption{Calculated fusion probabilities, $P_{\rm CN}(E^*,l=0)$,
for near-barrier collisions of heavy nuclei with $^{208}$Pb
target. CN excitation energies at the Bass barriers are shown by
the arrows. Experimental values of $P_{\rm CN}$ obtained in
\cite{Naik07} for the $^{50}$Ti+$^{208}$Pb fusion reaction are
shown by the rectangles.\label{pcne}}
\end{figure}

The main reason for low yields of evaporation residues in these
reactions is, however, a sharp decrease of the fusion probability
with increasing charge of the projectile. In Fig.~\ref{cold_cs}
the calculated capture, CN formation and evaporation residue (EvR)
cross sections of the $^{208}$Pb induced fusion reactions are
shown along with available experimental data on the yields of SH
elements (not all experimental points are displayed to simplify
the plot). The fusion probabilities $P_{\rm CN}$, calculated for
head-on collisions (which bring the main contribution to the EvR
cross sections), demonstrate a sharp energy dependence (see
Fig.~\ref{pcne}), found earlier in \cite{Zag01}. Recently the
decrease of the fusion probability at subbarrier energies was
confirmed experimentally for the fusion of $^{50}$Ti with
$^{208}$Pb \cite{Naik07}.

We found that the calculated energy dependence of the fusion
probability (shown in Fig.~\ref{pcne}) may be approximated by the
simple formula
\begin{equation}
P_{\rm CN}(E^*,l)=\frac{P_{\rm CN}^0}{1+exp[{\displaystyle
\frac{E^*_{\rm B}-E^*_{\rm int}(l)}{\Delta}}]}, \label{Pcn}
\end{equation}
which could be useful for a fast estimation of EvR cross sections
in the ``cold'' fusion reactions. Here $E^*_{\rm B}$ is the
excitation energy of CN at the center-of-mass beam energy equal to
the Bass barrier \cite{Bass80}. $E^*_{\rm B}$ are shown in
Fig.~\ref{pcne} by the arrows. $E^*_{\rm int}(l)=E_{\rm
c.m.}+Q-E_{\rm rot}(l)$ is the ``internal'' excitation energy
which defines also the damping of the shell correction to the
fission barrier of CN. $\Delta$ is the adjustable parameter of
about 4~MeV, and $P_{\rm CN}^0$ is the ``asymptotic''
(above-barrier) fusion probability dependent only on a combination
of colliding nuclei.

\begin{figure}[t]
\begin{center}
\includegraphics[width = 8.0 cm, bb=0 0 1028 798]{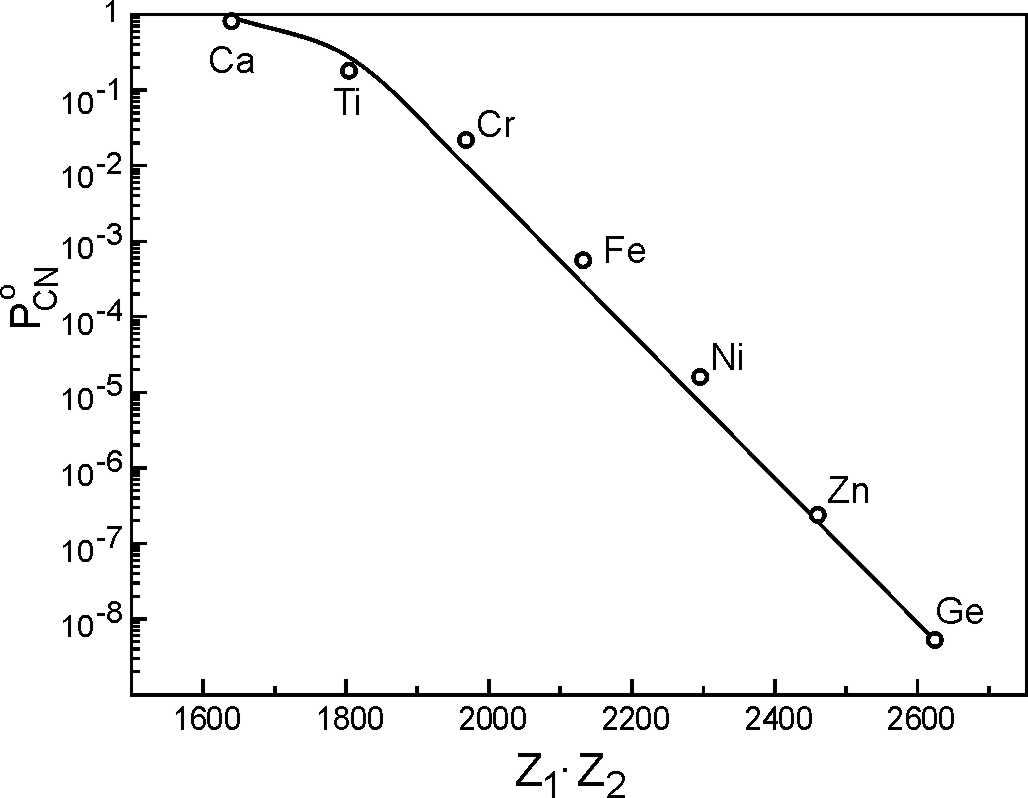} \end{center}
\caption{Above-barrier CN formation probability in the $^{208}$Pb
induced fusion reactions. Results of calculation are shown by the
circles, whereas the fitted curve corresponds to the expression
(\ref{Pcn0}).\label{pcn0}}
\end{figure}

The values of $P_{\rm CN}^0$ calculated at excitation energy
$E^*=40$~MeV (well above the barriers for the ``cold'' fusion
reactions) demonstrate rather simple behavior (almost linear in
logarithmic scale), monotonically decreasing with increase of
charge of CN and/or with increase of the product of $Z_1$ and
$Z_2$, see Fig.~\ref{pcn0}. This behavior could be also
approximated by very simple Fermi function
\begin{equation}
P_{\rm CN}^0 =\frac{1}{1+exp[\displaystyle \frac{Z_1Z_2 -
\zeta}{\tau}]}, \label{Pcn0}
\end{equation}
where $\zeta\approx 1760$ and $\tau\approx 45$ are just the fitted
parameters. Eq.(\ref{Pcn0}) is obviously valid only for the
``cold'' fusion reactions of heavy nuclei with the closed shell
targets $^{208}$Pb and $^{209}$Bi. Unfortunately we have not
enough experimental data to check this formula for other reactions
(or to derive more general expression for the fusion probability).

Two important remarks could be done after our analysis of the
``cold'' fusion reactions. The first is rather evident. There are
no reasons (in fusion or in survival probabilities) to slow down
the fast monotonic decrease of EvR cross sections with increasing
charge of SH nucleus synthesized in the ``cold'' fusion reaction.
The yield of 114 element in the 1n evaporation channel of the
$^{76}$Ge+$^{208}$Pb fusion reaction is only 0.06~pb. For 116 and
118 elements, synthesized in fusion reactions of $^{82}$Se and
$^{86}$Kr with lead target, we found only 0.004~pb and 0.0005~pb,
correspondingly, for 1n EvR cross sections (it is worth to note
that our results disagree with those obtained within the ``concept
of the dinuclear system'' \cite{Feng07}, which predicts the EvR
cross sections at the level of 0.1~pb for all these elements
including Z=120). As already mentioned, fusion reactions with
$^{208}$Pb or $^{209}$Bi targets lead to neutron deficient SH
nuclei with short half-lives, that may bring an additional
difficulty to their experimental detection at the available
separators.

The second conclusion is important for further experiments with
actinide targets. The experimental value of EvR cross section for
104 element in the $^{50}$Ti+$^{208}$Pb fusion reaction is two
orders of magnitude less as compared with the yield of 102 element
in the $^{48}$Ca+$^{208}$Pb reaction, see Fig.~\ref{cold_cs}. At
first sight, this fact makes the fusion reactions of titanium with
actinide targets (``hot'' fusion) much less encouraging as
compared to $^{48}$Ca fusion reactions. However, this sharp
decrease in the yield of the Rutherfordium isotopes is caused by
the two reasons. One order of magnitude loss in the EvR cross
section is due to the low survival probability of $^{258}$Rf
nucleus (the fission barrier is less by 0.4~MeV and neutron
separation energy is higher by 0.5~Mev as compared with
$^{256}$No, Fig.~\ref{Pcn}), whereas the fusion probability of
$^{50}$Ti with $^{208}$Pb at energies near and above the Coulomb
barrier is only one order of magnitude less than in the
$^{48}$Ca+$^{208}$Pb fusion reaction (see Fig.~\ref{pcne}). This
makes titanium beam quite promising for synthesis of SH nuclei in
fusion reactions with the actinide targets (see below).

\section{Hot fusion reactions}\label{Hot}

Fusion reactions of $^{48}$Ca with actinide targets lead to
formation of more neutron rich SH nuclei as compared to the
``cold'' fusion reactions. Their half-lives are several orders of
magnitude longer. For example, the half-life of the SH nucleus
$^{277}112$ synthesized in the ``cold'' fusion reaction
$^{70}$Zn+$^{208}$Pb \cite{Hoffman00,Morita} is about 1~ms,
whereas $T_{1/2}(^{285}112)\sim 34$~s \cite{Ogan07} (approaching
the ``island of stability''). On average, these SH nuclei have
higher fission barriers and lower neutron separation energies,
which give them a chance to survive in the neutron evaporation
cascade.

Unfortunately, weaker binding energies of the actinide nuclei lead
to rather high excitation energies of obtained CN (that is why
these reactions are named ``hot''). At beam energy close to the
Bass barrier the value of $E_{\rm CN}^*=E_{\rm c.m.}+B(Z_{\rm
CN},A_{\rm CN})-B(Z_1,A_1)-B(Z_2,A_2)$ ($B$ is the binding energy)
is usually higher than 30~MeV for almost all the combinations, and
at least 3 neutrons are to be evaporated to get a SH nucleus in
its ground state. The total survival probability of CN formed in
the ``hot'' fusion reaction (in the 3n and/or in the 4n channel)
is much less than 1n-survival probability in the ``cold'' fusion
reaction, $P_{\rm 3n}^{\rm ``hot''}(E^*\sim 35\,{\rm MeV}) <<
P_{\rm 1n}^{\rm ``cold''}(E^*\sim 15\,{\rm MeV})$.

On the other hand, for the more asymmetric ``hot'' combinations
the fusion probability is usually much higher as compared to the
``cold'' combinations leading to the same (but more neutron
deficient) elements. We calculated the capture, fusion and EvR
cross sections for the ``cold'' ($^{208}$Pb induced) and ``hot''
($^{48}$Ca induced) reactions leading to SH nuclei with $Z=102\div
118$ at the same excitation energies of the CN -- 15~MeV for the
``cold'' and 35~MeV for the ``hot'' combinations. Of course, the
beam energies, at which these CN excitations arise, are equal only
approximately to the corresponding Coulomb barriers and not all
them agree precisely with positions of maxima of EvR cross
sections. However, some general regularities can be found from
these calculations.

\begin{figure}[ht]
\begin{center}
\includegraphics[width = 8.0 cm, bb=0 0 1702 1848]{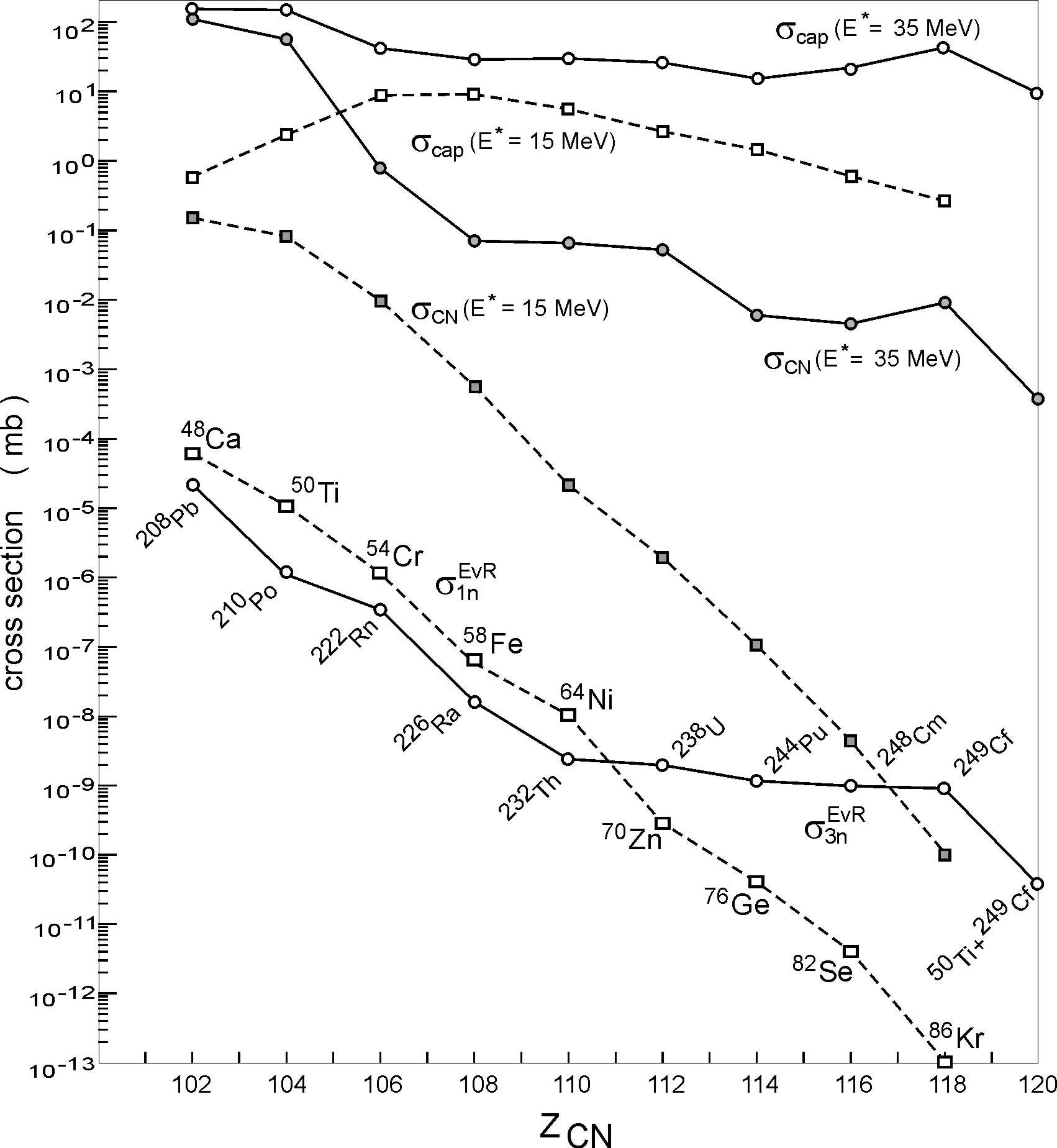} \end{center}
\caption{Calculated capture, fusion and evaporation residue cross
sections in the ``cold'' $^{208}$Pb induced (rectangles joined by
dashed lines, projectiles are shown) and ``hot'' $^{48}$Ca induced
(circles joined by solid lines, targets are shown) fusion
reactions. The cross sections are calculated at beam energies
corresponding to 15~MeV (``cold'' fusion, 1n channel) and 35~MeV
(``hot'' fusion, 3n channel) excitation energies of the compound
nuclei.\label{cold_hot}}
\end{figure}

The results of our calculations are shown in Fig.~\ref{cold_hot}.
As can be seen, the capture cross sections are about one order of
magnitude larger for the ``hot'' combinations. This is because the
$E^*=15$~MeV corresponds to the incident energies somewhat below
the Bass barriers of the ``cold'' combinations. Slow decrease of
$\sigma_{\rm cap}$ for the ``cold'' combinations at $Z_{\rm
CN}>108$ is caused by gradual shallowing of the potential pocket
(decreasing value of $l_{\rm crit}$). Larger value of $\sigma_{\rm
cap}$ for the $^{48}$Ca+$^{249}$Cf combination is conditioned by a
``colder'' character of this reaction -- the excitation energy of
CN at the Bass barrier beam energy is only 28~MeV for this
reaction (i.e., $E^* = 35$~MeV corresponds here to above barrier
initial energy).

The fusion probability for the ``cold'' combinations decreases
very fast with increasing charge of the projectile and, in spite
of evaporation of only one neutron, at $Z_{\rm CN}\ge 112$ the EvR
cross sections become less than in ``hot'' fusion reactions.
Increasing survival probability of SH nuclei with $Z = 114, 116$
synthesized in $^{48}$Ca induced fusion reactions as compared to
$Z = 110, 112$ is due to the increase of the shell corrections to
the fission barriers of these nuclei caused by approaching the
closed shells predicted by the macro-microscopical model (see
Table~\ref{bfSH}).

\begin{table}[h]
\caption{Fission barriers (macroscopical part and shell
correction) and neutron separation energies (MeV) of CN produced
in the $^{48}$Ca fusion reactions with $^{232}$Th, $^{238}$U,
$^{244}$Pu, $^{248}$Cm and $^{249}$Cf targets \cite{Moller95}. The
last column shows the excitations of CN at the Bass barrier
incident energies. \label{bfSH}}
\begin{tabular}{|c|c|c|c|c|c|}
\hline
CN & B$_{\rm LD}$ & Sh.Corr. &  $B_{\rm fis}$ & $E_n^{\rm sep}$ & $E^*$(Bass)\\[2pt]
\hline
$^{280}$110 & 0.21 & 4.76 & 5.0  & 7.0 & 32  \\[2pt]
$^{286}$112 & 0.10 & 6.64 & 6.7  & 7.1 & 33  \\[2pt]
$^{292}$114 & 0.04 & 8.89 & 8.9  & 7.0 & 34  \\[2pt]
$^{296}$116 & 0.01 & 8.58 & 8.6  & 6.7 & 32  \\[2pt]
$^{297}$118 & 0.00 & 8.27 & 8.3  & 6.2 & 28  \\[2pt]
\hline
\end{tabular}
\end{table}

\begin{figure}[ht]
\begin{center}
\includegraphics[width = 8.0 cm, bb=0 0 1051 1774]{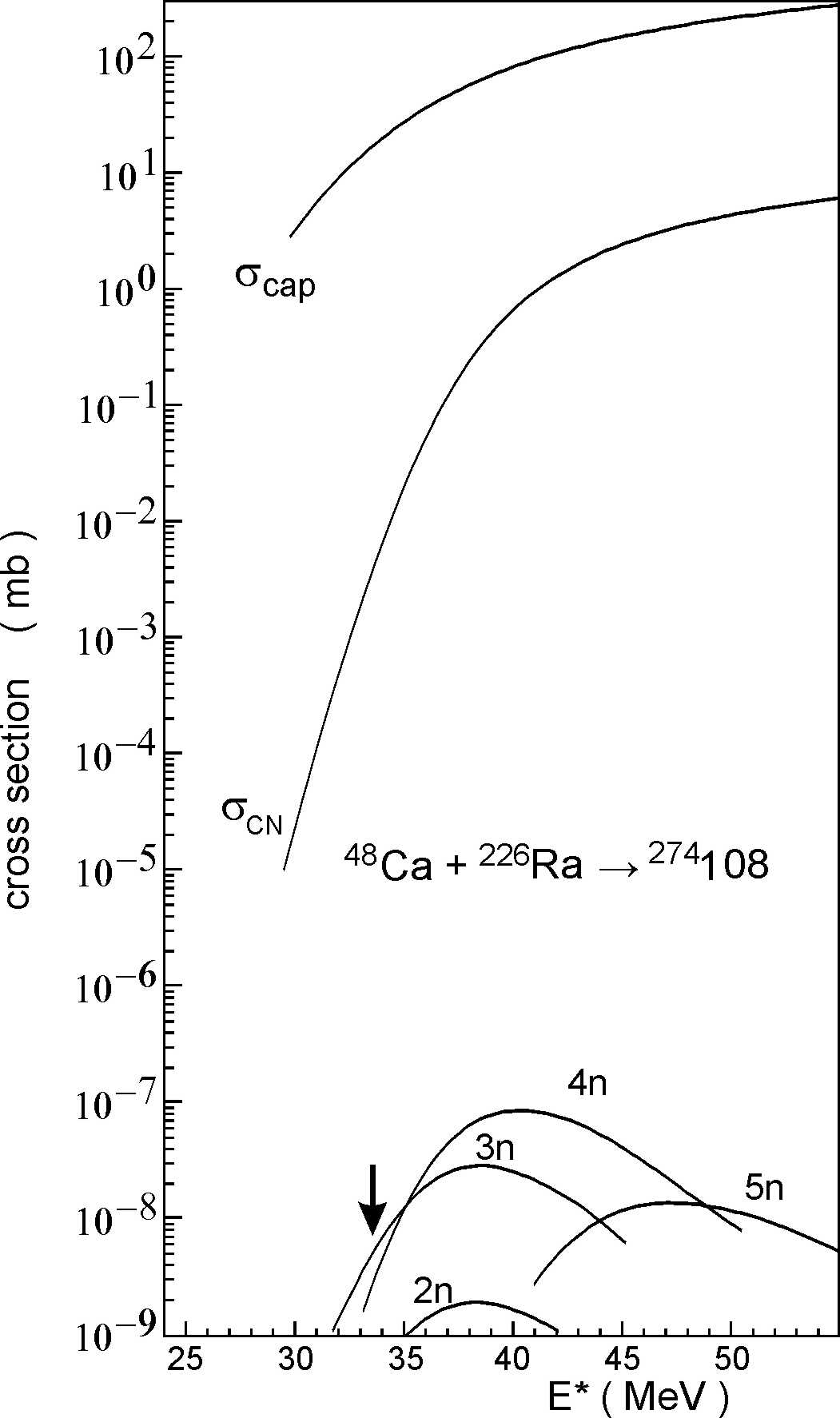} \end{center}
\caption{Calculated capture, fusion and evaporation residue (2n,
3n, 4n and 5n channels) cross sections in the $^{48}$Ca+$^{226}$Ra
fusion reaction. The arrow indicates the Bass barrier. \label{Ra}}
\end{figure}

In the experimental data on the ``hot'' fusion reactions induced
by $^{48}$Ca there is unexplored gap between the elements 102
($^{208}$Pb target) and 112 ($^{238}$U target). For deeper
understanding of the mechanisms of SH element formation, an
additional point in this region (where the cross section falls
down by four orders of magnitude) is extremely desirable. We found
that the neutron rich isotopes of Hassium (Z=108) could be
produced in the $^{48}$Ca+$^{226}$Ra fusion reaction with rather
large cross sections, Fig.~\ref{Ra}. In such experiment one should
worry about utilization of $^{222}$Rn (decaying finally to rather
long-lived $^{210}$Po), however, $^{226}$Ra target was already
used in the past. Simultaneous measurement of the capture cross
section could be also rather useful for subsequent theoretical
analysis. Note, that our estimation of the EvR cross sections in
this reaction is rather close to those obtained in \cite{Liu06}.

\begin{figure}[t]
\begin{center}
\includegraphics[width = 8.0 cm, bb=0 0 1205 642]{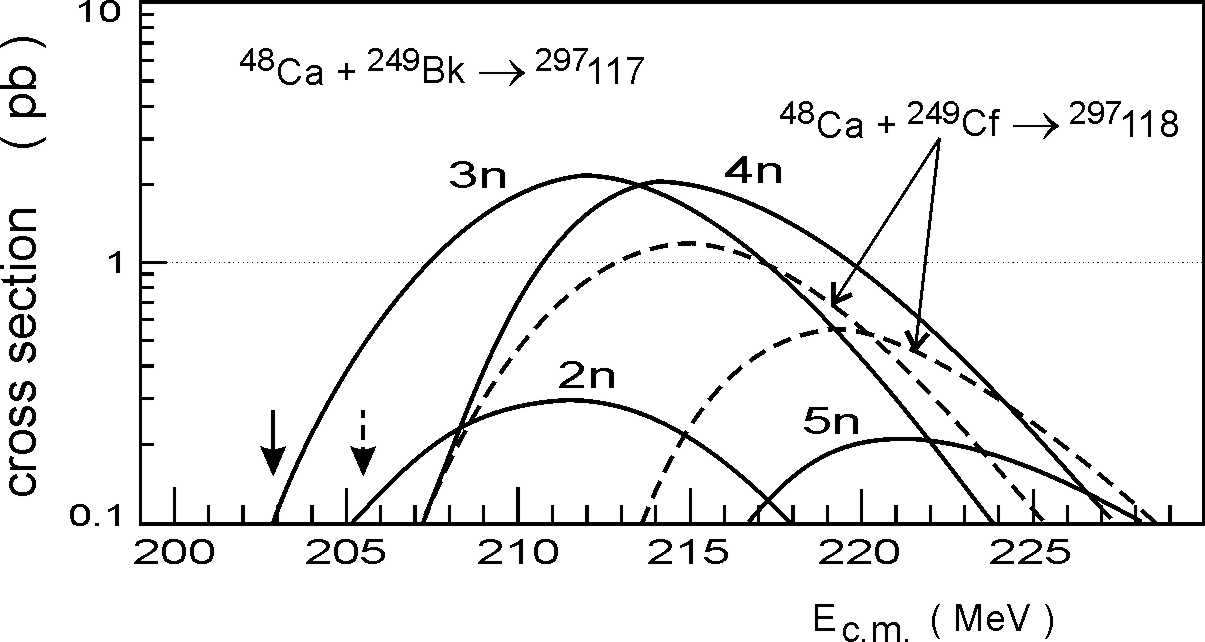} \end{center}
\caption{Cross sections for production of the element 117 in the
$^{48}$Ca+$^{249}$Bk fusion reaction (solid curves, 2n, 3n, 4n and
5n evaporation channels). For comparison the EvR cross sections in
3n and 4n channels of the $^{48}$Ca+$^{249}$Cf fusion reaction are
shown by the dashed curves. The arrows indicate the corresponding
Bass barriers.\label{117}}
\end{figure}

In the series of SH elements synthesized in the $^{48}$Ca induced
fusion reactions \cite{Ogan07} one element, Z=117, is still
``skipped''. The element 117 may be synthesized with rather large
cross section in the $^{48}$Ca+$^{249}$Bk fusion reaction, if one
manages to prepare a short-living (330 days) berkelium target. The
calculated EvR cross sections of this reaction are shown in
Fig.~\ref{117}. It is important that the successive nuclei
($^{289,290}$115, $^{285,286}$113, $^{281,282}$111,
$^{277,278}$109, and so on) appearing in the $\alpha$-decay chains
of $^{293,294}$117 are assumed to have rather long half-lives to
be detected and studied in the chemical experiment, that makes the
$^{48}$Ca+$^{249}$Bk fusion reaction quite attractive. Also the
berkelium target may be used for synthesis of the element 119 in
fusion reaction with the titanium beam (see below).

\begin{figure}[t]
\begin{center}
\includegraphics[width = 8.0 cm, bb=0 0 1225 1216]{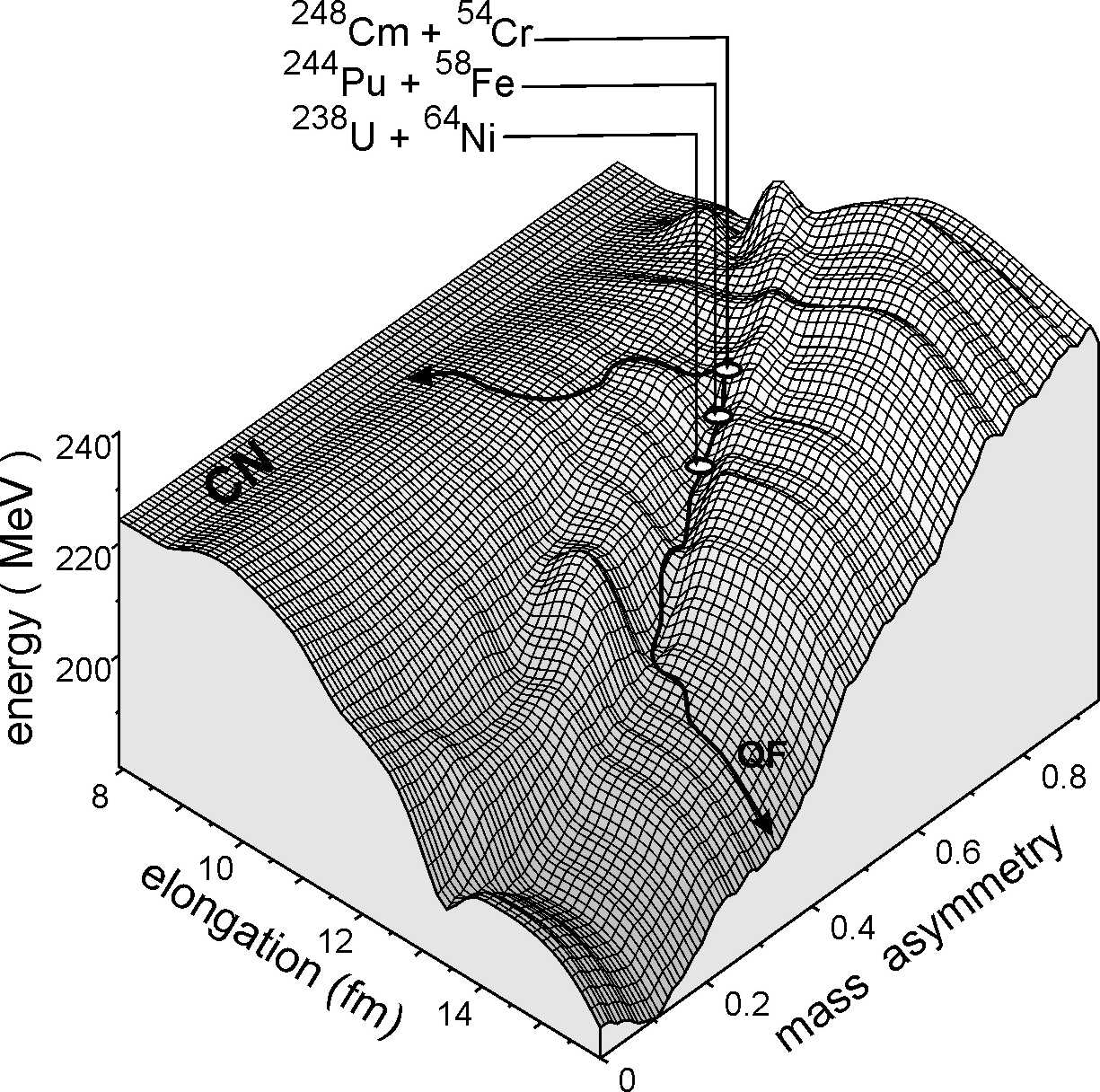} \end{center}
\caption{Potential energy surface for the nuclear system
consisting of 120 protons and 182 neutrons
(elongation--mass-asymmetry plot at fixed dynamic deformation
$\beta_2=0.2$). Injection configurations (contact points) for the
$^{54}$Cr+$^{248}$Cm, $^{58}$Fe+$^{244}$Pu and $^{64}$Ni+$^{238}$U
fusion reactions are shown by the circles. Thick curves with
arrows shows schematically quasi-fission and fusion (CN formation)
trajectories.\label{120}}
\end{figure}

As mentioned above, $^{249}$Cf ($T_{1/2}=351$~y) is the heaviest
available target which may be used in experiment. Thus, to get SH
elements with $Z>118$ in fusion reactions we should proceed to
heavier than $^{48}$Ca projectiles. Most neutron-rich isotopes of
120-th element may be synthesized in the three different fusion
reactions $^{54}$Cr+$^{248}$Cm, $^{58}$Fe+$^{244}$Pu and
$^{64}$Ni+$^{238}$U leading to the same SH nucleus $^{302}120$
with neutron number near to the predicted closed shell N=184.
These three combinations are not of equal worth. In Fig.~\ref{120}
the potential energy surface for the nuclear system consisting of
120 protons and 182 neutrons is shown in the
``elongation--mass-asymmetry'' space at fixed value of dynamic
deformation $\beta_2=0.2$. One can see that the contact
configuration of the more symmetric $^{64}$Ni+$^{238}$U
combination is located lower in the valley leading the nuclear
system to the dominating quasi-fission channels.

\begin{figure}[ht]
\begin{center}
\includegraphics[width = 8.0 cm, bb=0 0 1087 527]{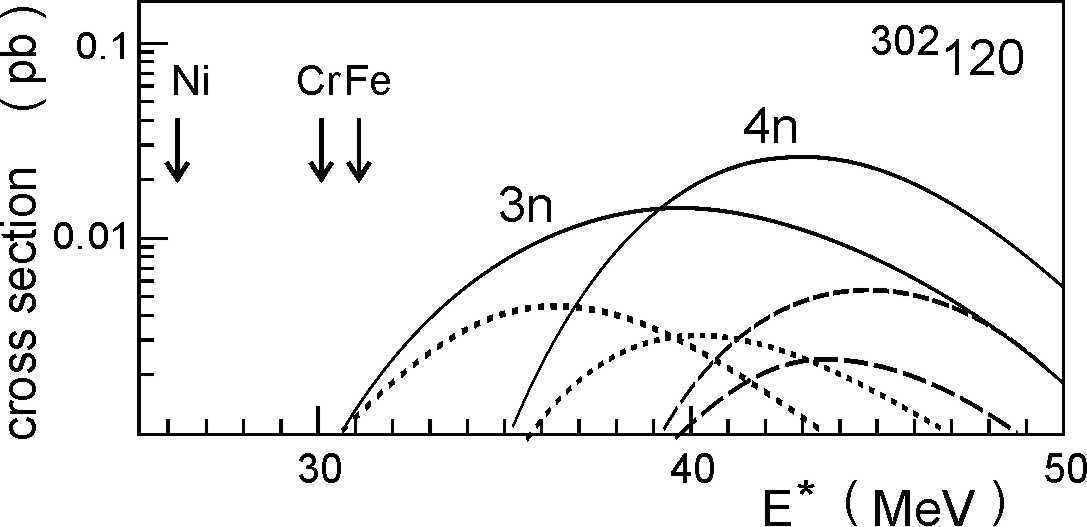} \end{center}
\caption{Excitation functions for production of the Z=120 element
in 3n and 4n evaporation channels of the $^{54}$Cr+$^{248}$Cm
(solid curves), $^{58}$Fe+$^{244}$Pu (dashed) and
$^{64}$Ni+$^{238}$U (dotted) fusion reactions. The corresponding
Bass barriers are shown by the arrows.\label{120crs}}
\end{figure}

As a result the estimated EvR cross sections for more symmetric
$^{58}$Fe+$^{244}$Pu and $^{64}$Ni+$^{238}$U reactions are lower
as compared to the less symmetric $^{54}$Cr+$^{248}$Cm combination
(see Fig.~\ref{120crs}). Some gain for $^{64}$Ni+$^{238}$U comes
from the ``colder'' character of this reaction -- the excitation
of CN at the Bass barrier incident energy for this combination,
$E^*_{\rm CN}=26$~MeV, is much lower than for two others (see
arrows in Fig.~\ref{120crs}). Note, that 3n and 4n evaporation
residues of the $^{302}$120 nucleus will decay over the known
isotopes of 112~$\div$~118 elements \cite{Ogan07}. This
significantly simplifies their identification. However, the
$Q$-value of the first $\alpha$-particle emitted from the element
120 should be rather high (about 13~MeV) and the half-life of this
element might be rather short. If it is comparable with the time
of flight of the recoil nucleus through a separator (about 1~$\mu
s$), then an additional difficulty appears in detection of this
element.

When calculating survival probability we used the fission barriers
of SH nuclei predicted by the macro-microscopical model
\cite{Moller95}, which gives much lower fission barrier for
$^{302}$120 nucleus as compared to $^{296}$116. On the other hand,
the full microscopic models based on the self-consistent
Hartree--Fock calculations \cite{Burv04} predict much higher
fission barriers for the nucleus $^{302}$120 (up to 10~MeV) if the
Skyrme forces are used (though these predictions are not
unambiguous and depend strongly on chosen nucleon-nucleon forces).
This means that the estimated 3n and 4n EvR cross sections in the
fusion reactions considered above could be, in principle, higher
than those shown in Fig.~\ref{120crs}. This fact, however,
influences neither the positions of the maxima of the excitation
functions nor the conclusion about the advantage of the
$^{54}$Cr+$^{248}$Cm fusion reaction as compared to
$^{64}$Ni+$^{238}$U.

\begin{figure}[h]
\begin{center}
\includegraphics[width = 6.6 cm, bb=0 0 1123 3085]{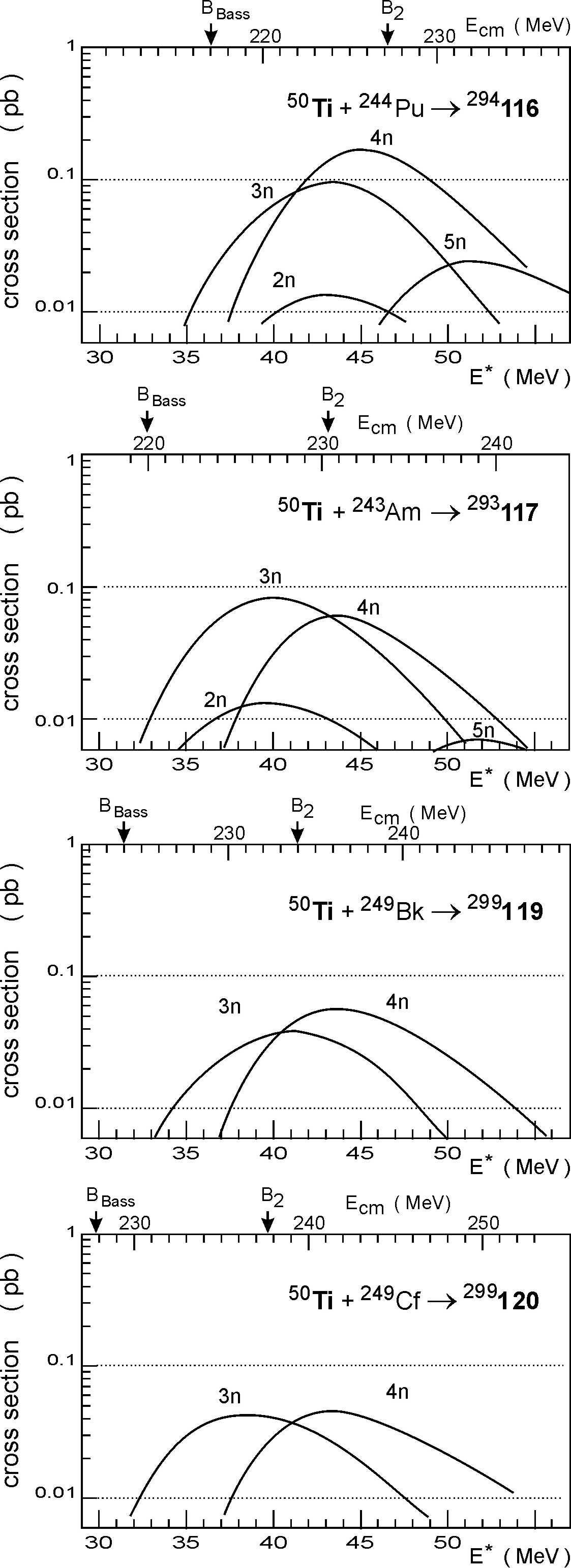} \end{center}
\caption{Excitation functions of $^{50}$Ti induced synthesis of
116, 117, 119 and 120 elements. The arrows indicate the positions
of the corresponding Bass barriers and the Coulomb barriers of
side-by-side oriented nuclei.\label{Ti}}
\end{figure}

Strong dependence of the calculated EvR cross sections for the
production of 120 element on mass-asymmetry in the entrance
channel (along with their low values for all the reactions
considered above) makes the nearest to $^{48}$Ca projectile,
$^{50}$Ti, most promising for further synthesis of SH nuclei. Of
course, the use of the titanium beam instead of $^{48}$Ca also
decreases the yield of SH nuclei mainly due to a worse fusion
probability. The calculated excitation functions for synthesis of
116, 117, 119 and 120 SH elements in the fusion reactions of
$^{50}$Ti with $^{244}$Pu, $^{243}$Am, $^{249}$Bk and $^{249}$Cf
targets are shown in Fig.~\ref{Ti}.

The orientation effects are known to play an important role in
fusion reactions of statically deformed heavy nuclei
~\cite{Zag04a,Zag04b,ZG07}. The fusion probability (formation of
CN) was found to be strongly suppressed for more elongated
nose-to-nose initial orientations ~\cite{ZG07}. As a result the
preferable beam energies for synthesis of SH elements in the
``hot'' fusion reactions are shifted to values which are several
MeV higher then the corresponding Bass barriers (calculated for
spherical nuclei). As can be seen from Fig.~\ref{Ti}, the
estimated EvR cross sections for 117, 119 and 120 SH elements
synthesized in the $^{50}$Ti induced reactions are quite reachable
at available experimental setups, though one needs longer time of
irradiation as compared with $^{48}$Ca fusion reactions.

\section{Mass symmetric fusion reactions}\label{Symm}

The use of the accelerated neutron-rich fission fragments is one
of the widely discussed speculative methods for the production of
SH elements in the region of the ``island of stability''. For
example, in the $^{132}$Sn+$^{176}$Yb fusion reaction we may
synthesize $^{308}$120, which (after a few neutron evaporations
and $\alpha$-decays) may reach a center of the ``island of
stability''. Several projects in the world are now realizing to
get the beams of neutron rich fission fragments. The question is
how intensive should be such beams to produce SH nuclei. Evidently
the answer depends on the values of the corresponding cross
sections. Unfortunately, there are almost no experimental data on
fusion reactions in mass-symmetric nuclear combinations.

Experimental data on symmetric fusion reactions
$^{100}$Mo+$^{100}$Mo, $^{100}$Mo+$^{110}$Pa and
$^{110}$Pa+$^{110}$Pa \cite{MoPa} show that the fusion probability
sharply decreases with increasing mass and charge of colliding
nuclei. However, the last studied reactions of such kind,
$^{110}$Pa+$^{110}$Pa, is still far from a combination leading to
a SH compound nucleus. This means that further experimental study
of such reactions is quite urgent.

The choice of the colliding nuclei is also important. In this
connection the $^{136}$Xe+$^{136}$Xe fusion reaction looks very
promising for experimental study \cite{Ogan06b}, because the
formed CN, $^{272}$Hs, should undergo just to symmetric fission.
It means that two colliding $^{136}$Xe nuclei are very close to
the nascent fission fragments of $^{272}$Hs in the region of the
saddle point, and their fusion should really reflect a fusion
process of two fission fragments.

\begin{figure}[ht]
\begin{center}
\includegraphics[width = 8.0 cm, bb=0 0 1326 1260]{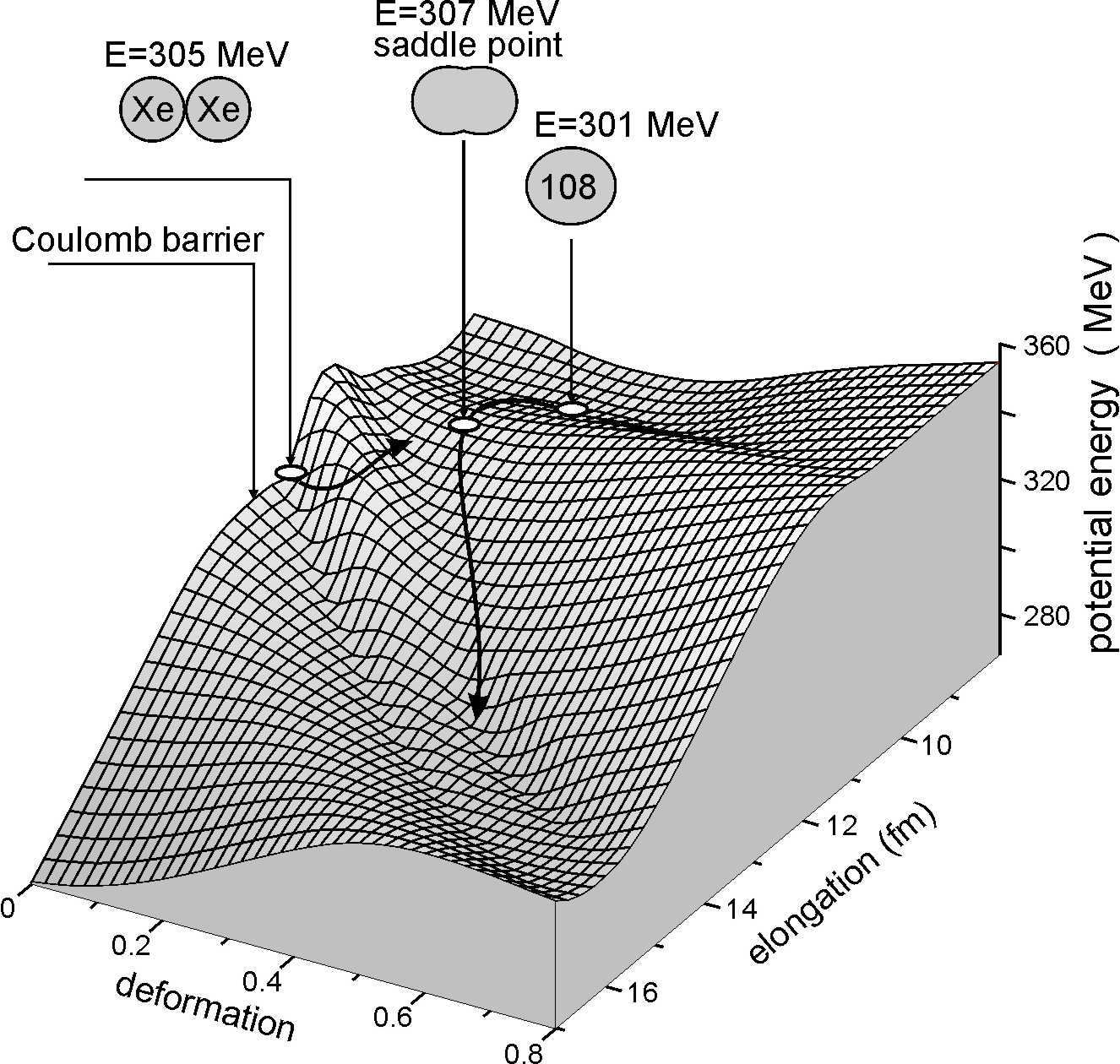} \end{center}
\caption{Adiabatic potential energy of the $^{272}$108 nuclear
system at zero mass asymmetry ($^{136}$Xe+$^{136}$Xe configuration
in asymptotic region) in the ``elongation--deformation'' space.
The curves with arrows show the fission and fusion paths. The
circles show positions of CN, saddle point and contact
configuration of two spherical Xe nuclei.\label{xexepot}}
\end{figure}

The calculated within the two-center shell model adiabatic
potential energy surface of the nuclear system consisting of 108
protons and 164 neutrons is shown in Fig.~\ref{xexepot} as a
function of elongation (distance between the centers) and
deformation of the fragments at zero mass asymmetry, which
correspond to two Xe nuclei in the entrance and exit channel. The
energy scale is chosen in such a way that zero energy corresponds
to two $^{136}$Xe nuclei in their ground states at infinite
distance. The contact configuration of two spherical Xe nuclei is
located very close (in energy and in configuration space) to the
saddle point of CN (note that it is located behind the Coulomb
barrier, though there is no pronounced potential pocket). This
fusion reaction is extremely ``cold'', the excitation energy of
the CN at the Bass barrier beam energy is only 5~MeV. One may
expect that after contact these nuclei may overcome the inner
barrier due to fluctuations of collective degrees of freedom and
thus reach the saddle configuration. After that they fuse (form
CN) with 50\% probability.

\begin{figure}[ht]
\begin{center}
\includegraphics[width = 8.0 cm, bb=0 0 1047 807]{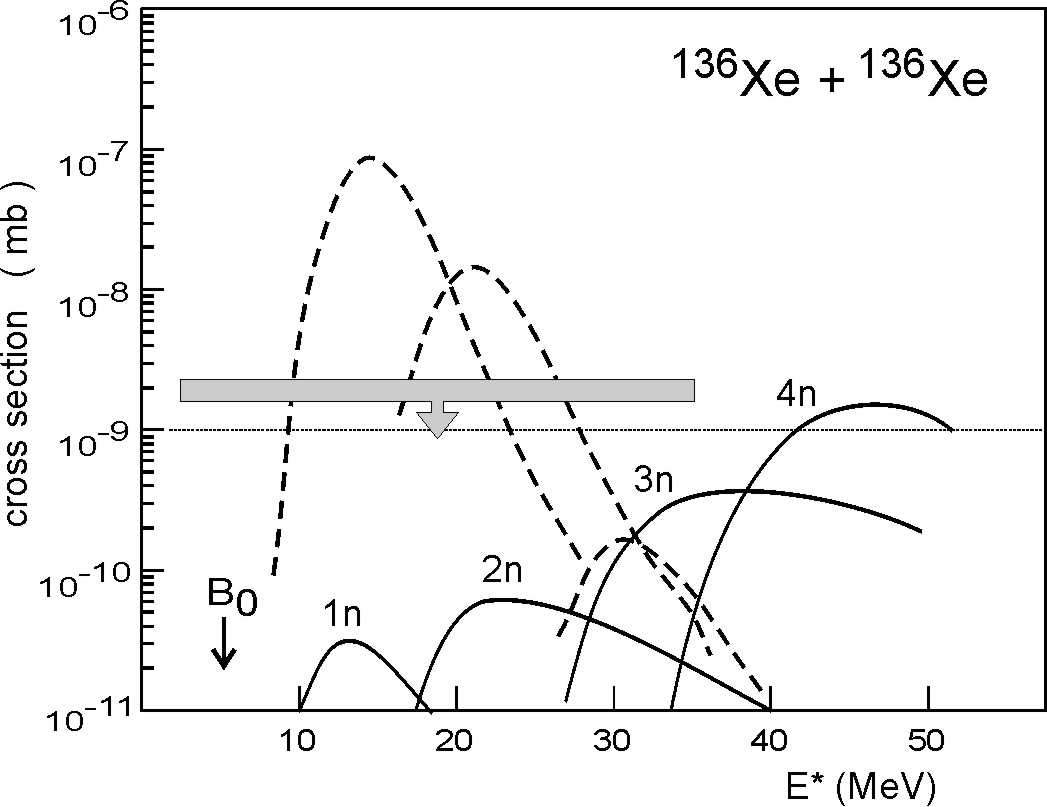} \end{center}
\caption{Evaporation residue cross sections in the
$^{136}$Xe+$^{136}$Xe fusion reactions. Solid lines show our
predictions \cite{ZGnp07}, whereas the dashed curves are the
predictions taken from Ref.~\cite{Sw07}. Gray bar shows upper
limit of the experimental EvR cross sections in this reaction
\cite{Eremin}.\label{xexecrs}}
\end{figure}

However the potential energy decreases very fast with increasing
deformations of the touching nuclei and drives the nuclear system
to the fission valley (see Fig.~\ref{xexepot}). As a result, the
calculated fusion probability is very low and, in spite of rather
high fission barriers of the hassium isotopes in the region of
$A\sim 270$ ($\sim 6$~MeV \cite{Moller95}), the EvR cross sections
were found to be very low \cite{ZGnp07}, see Fig.~\ref{xexecrs}.
They are much less than the yield of $^{265}$Hs synthesized in the
more asymmetric $^{58}$Fe+$^{208}$Pb fusion reaction
(Fig.~\ref{cold_cs}). It is worthy to note that the prediction of
the EvR cross section for the 1n channel in the
$^{136}$Xe+$^{136}$Xe fusion reaction, obtained within the
so-called ``diffusion model'' \cite{Sw05}, exceeds our result by
three orders of magnitude. This fact reflects significant
difficulties appearing in the calculation of the fusion
probability in such reactions.

Experiment on the synthesis of hassium isotopes in the
$^{136}$Xe+$^{136}$Xe fusion reaction was performed recently in
Dubna, and no one event was detected at the level of about 2~pb
\cite{Eremin}. Thus, we may conclude that for the widely discussed
future experiments on synthesis of SH nuclei in the fusion
reactions with accelerated fission fragments one needs to get a
beam intensity not lower than $10^{13}$~pps (comparable or greater
than intensities of available stable beams of heavy ions). Since
the experimental values of the EvR cross sections in such
reactions are still unknown, attempts to synthesize a SH element
in the fusion reaction of two heavy more or less equal in masses
nuclei (Xe+Xe or Sn+Xe) should be continued.

\section{Radioactive ion beams}\label{RIB}

Recently many speculations also appeared on the use of radioactive
beams for synthesis and study of new elements and isotopes. A
rather complete list of references as well as a review on this
problem can be found in the paper of Loveland \cite{Lov07}.

\begin{figure}[ht]
\begin{center}
\includegraphics[width = 8.0 cm, bb=0 0 2010 1742]{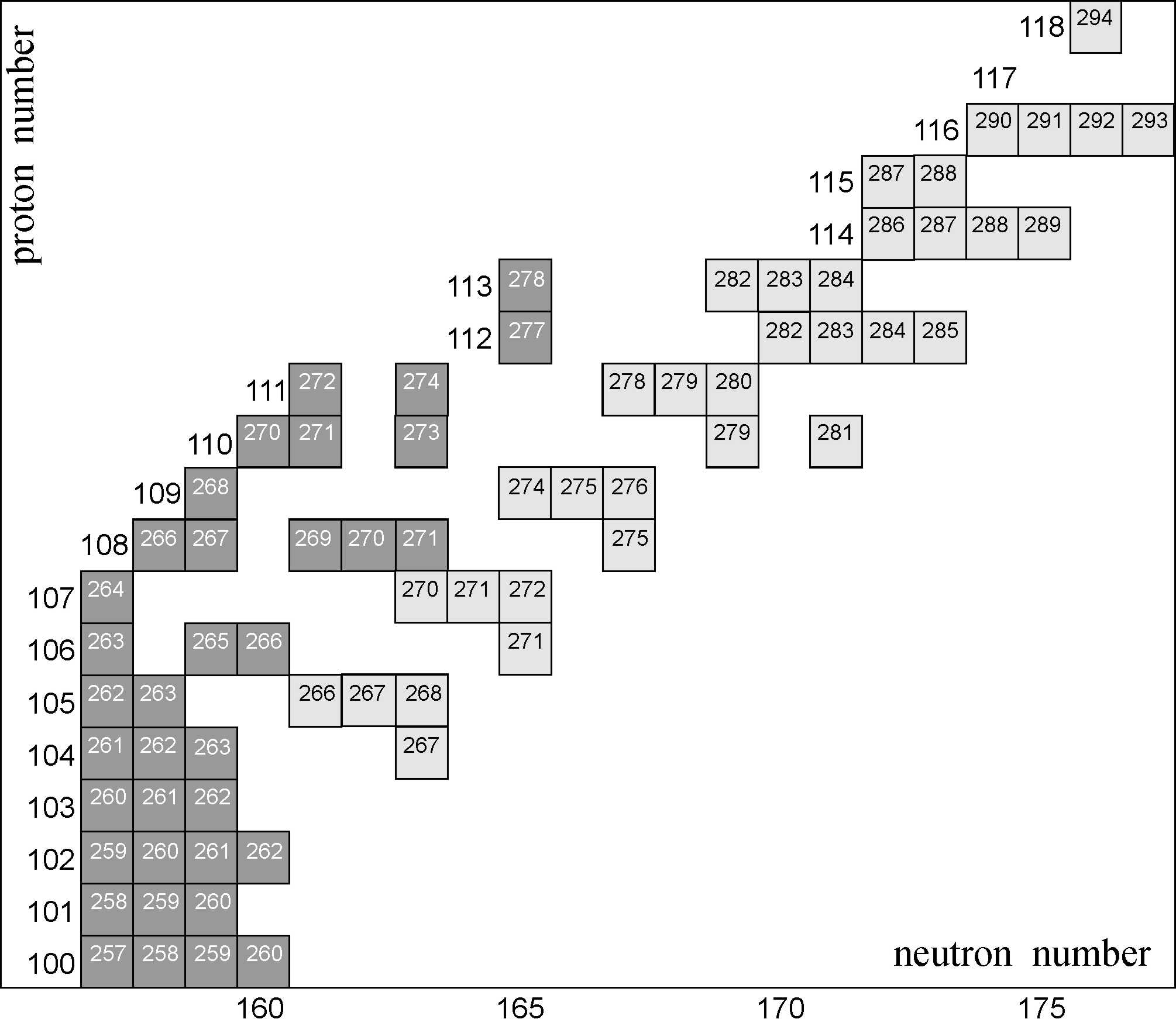} \end{center}
\caption{Upper part of the nuclear map. Isotopes synthesised in
the $^{48}$Ca induced fusion reactions are shown by the light gray
rectangles.\label{mapup}}
\end{figure}

As shown above, the use of accelerated fission fragments for the
production of SH nuclei in symmetric fusion reactions is less
encouraging and needs beam intensities at the hardly reachable
level of $10^{13}$~pps or higher. In our opinion, they are the
lighter radioactive beams which could be quite useful to solve the
two important problems. As can be seen from Fig.~\ref{mapup} there
is some gap between the SH nuclei produced in the ``hot'' fusion
reactions with $^{48}$Ca and the mainland. This gap hinders
obtaining a clear view on the properties of SH nuclei in this
region (in particular, positions of closed shells and sub-shells).
There are no combinations of stable nuclei to fill this gap in
fusion reactions, while the use of radioactive projectiles may
help to do this.

The second problem, which may be solved with the radioactive
beams, is obtaining much more neutron rich transfermium isotopes.
It is extremely important for two reasons. First, as we know from
experiment, the addition of only 8 neutrons to nucleus
$^{277}112(T_{1/2}=0.7$~ms) increases its half-life by almost 5
orders of magnitude -- $T_{1/2}(^{285}112)=34$~s -- testifying the
approach of the ``island of stability''. How far is it? How long
could be half-lives of SH nuclei at this island? To answer these
questions we need to add more and more neutrons. Second, somewhere
in the region of Z$\sim$100 and N$\sim$170 the r-process of
nucleosynthesis should be terminated by neutron-induced or
$\beta$-delayed fission. This region of nuclei, however, is
absolutely unknown and only theoretical estimations of nuclear
properties (rather unreliable for neutron rich isotopes) are
presently used in different astrophysical scenarios.

Contrary to a common opinion, neutron excess itself does not
increase very much the EvR cross sections in fusion reactions of
neutron rich radioactive nuclei. The neutron excess decreases just
a little the height of the Coulomb barrier due to the small
increase in the radius of neutron rich projectile. Neutron
transfer with positive $Q$-value may really increase the
sub-barrier fusion probability by several orders of magnitude due
to ``sequential fusion mechanism'' \cite{Zag03,ZSG07}. However,
this mechanism does not increase noticeably the fusion probability
at near-barrier incident energies, where the EvR cross sections
are maximal (see above).

\begin{figure}[ht]
\begin{center}
\includegraphics[width = 8.0 cm, bb=0 0 1332 610]{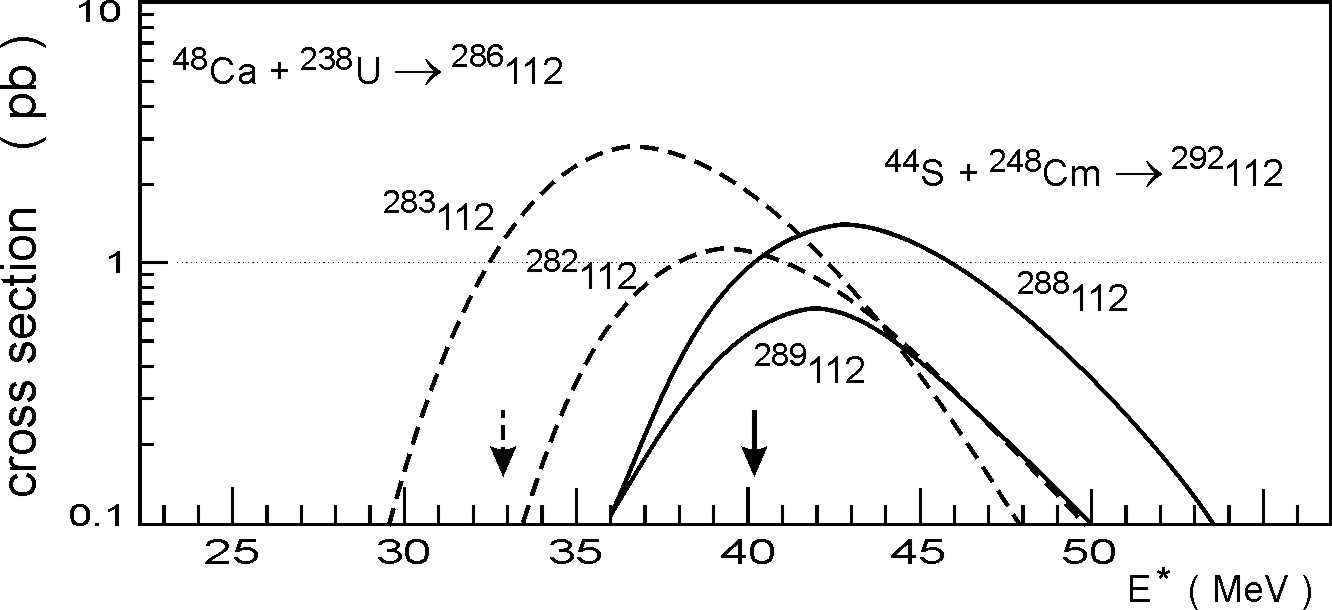} \end{center}
\caption{Excitation functions for the synthesis of the isotopes of
the element 112 in 3n and 4n evaporation channels of the
$^{48}$Ca+$^{238}$U (A=282 and A=283, dashed curves) and
$^{44}$S+$^{248}$Cm (A=288 and  A=289, solid curves) fusion
reactions. Arrows indicate the corresponding Bass barriers for the
two reactions.\label{112}}
\end{figure}

Fig.~\ref{112} shows the EvR cross sections for the
$^{44}$S+$^{248}$Cm fusion reaction, in which the isotopes of the
element 112 with six more neutrons (as compared with the
$^{48}$Ca+$^{238}$U reaction) could be synthesized. The calculated
one-picobarn cross sections mean that the beam intensity of
sulfur-44 (which may be produced, for example, by 4p stripping
from $^{48}$Ca) should be no less than $10^{12}$~pps to synthesize
these extremely neutron rich isotopes.

In utmost mass-asymmetric fusion reactions (with lighter than neon
projectiles) there is no suppression of CN formation: after
contact colliding nuclei form CN with almost unit probability,
$P_{\rm CN}\approx 1$. This significantly increases the EvR cross
sections in such reactions and, in spite of the rather difficult
production of light radioactive nuclei with significant neutron
excess, they could be used for the study of neutron rich
transfermium nuclei.

\begin{figure}[ht]
\begin{center}
\includegraphics[width = 8.0 cm, bb=0 0 1603 1046]{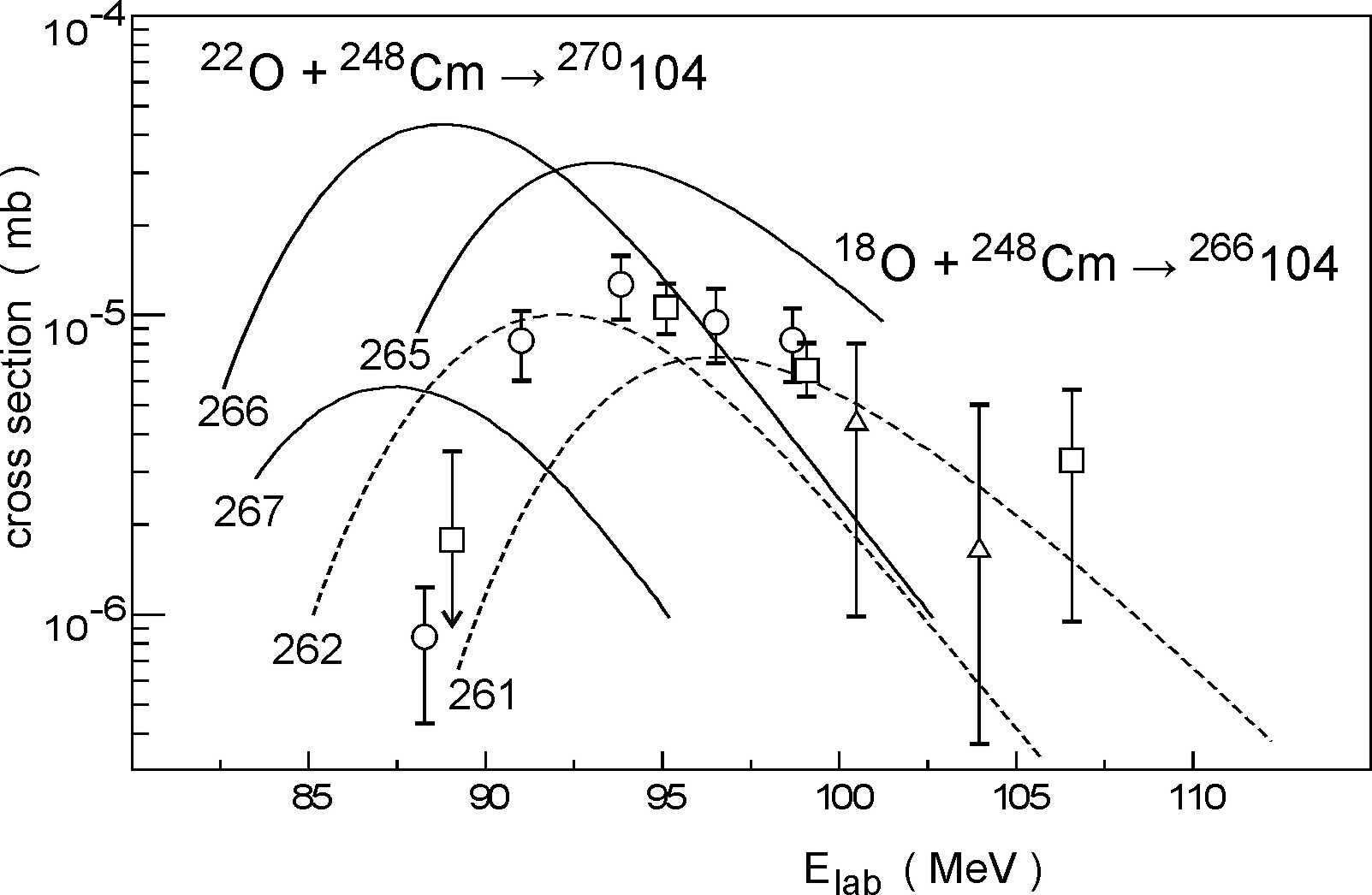} \end{center}
\caption{Excitation functions for synthesis of Rutherfordium
isotopes in the $^{18}$O+$^{248}$Cm (A=261 and A=262, dashed
curves) and $^{22}$O+$^{248}$Cm (A=265, A=266 and A=267, solid
curves) fusion reactions. Experimental data for the
$^{248}$Cm($^{18}$O,5n)$^{261}$Rf reaction are from \cite{Silva73}
(rectangles), \cite{Dressler99} (triangles) and \cite{Nagame02}
(circles).\label{104}}
\end{figure}

New heavy isotopes of Rutherfordium (up to $^{267}$104) might be
obtained in the $^{22}$O+$^{248}$Cm fusion reaction. The EvR cross
sections in this reaction (shown in Fig.~\ref{104}) are rather
large and the beam intensity of $^{22}$O at the level of
$10^{8}$~pps is sufficient to detect one decay event per week.
Note that the reaction $^{22}$O+$^{248}$Cm is 3~MeV ``colder'' as
compared to $^{18}$O+$^{248}$Cm ($E^*(\rm Bass)=41$ and 44~MeV,
respectively) that allows one to measure even the 3n evaporation
channel leading to $^{267}$104 (see Fig.~\ref{104}). Half-lives of
the heavy Rutherfordium isotopes (A$>$263) should be rather long
to use chemical methods for their identification.

\section{Multi-nucleon transfer reactions}\label{DIP}

The use of multi-nucleon transfer from heavy-ion projectile to an
actinide target nucleus for the production of new nuclear species
in the transuranium region has a long history. Light (carbon
\cite{Hahn74}, oxygen and neon \cite{Lee82}), medium (calcium
\cite{Hulet77,Turler92}, krypton and xenon \cite{Moody86,Welch87})
and very heavy ($^{238}$U \cite{Schadel78,Schadel82}) projectiles
were used and heavy actinides (up to Mendelevium) have been
produced in these reactions. The cross sections were found to
decrease very rapidly with increasing transferred mass and atomic
number of surviving target-like fragments. The level of 0.1~$\mu$b
was reached for chemically separated Md isotopes \cite{Schadel82}.

These experiments seem to give not so great chances for production
of new SH nuclei. However, there are experimental evidences that
the nuclear shell structure may strongly influence the nucleon
flow in the low-energy damped collisions of heavy ions. For
example, in $^{238}$U-induced reactions on $^{110}$Pd at about
6~MeV/u bombarding energy an enhanced proton flow along the
neutron shells $N_1=82$ and $N_2=126$ (reached almost
simultaneously in target-like and projectile-like fragments) was
observed in the distribution of binary reaction products
\cite{Mayer85}.

The idea to take advantage of the shell effects for the production
of SH nuclei in the multi-nucleon transfer processes of low-energy
heavy ion collisions was proposed in \cite{ZOIG06}. The shell
effects are known to play an important role in fusion of heavy
ions with actinide targets driving the nuclear system to the
quasi-fission channels (into the deep lead and tin valleys) and,
thus, decreasing the fusion probability. On the contrary, in the
transfer reactions the same effects may lead to enhanced yield of
SH nuclei. It may occur if one of heavy colliding nuclei, say
$^{238}$U, gives away nucleons approaching to double magic
$^{208}$Pb nucleus, whereas another one, say $^{248}$Cm, accepts
these nucleons becoming superheavy in the exit channel -- the so
called ``inverse'' (anti-symmetrizing) quasi-fission process.

\begin{figure}[t]
\begin{center}
\includegraphics[width = 8.0 cm, bb=0 0 1857 1351]{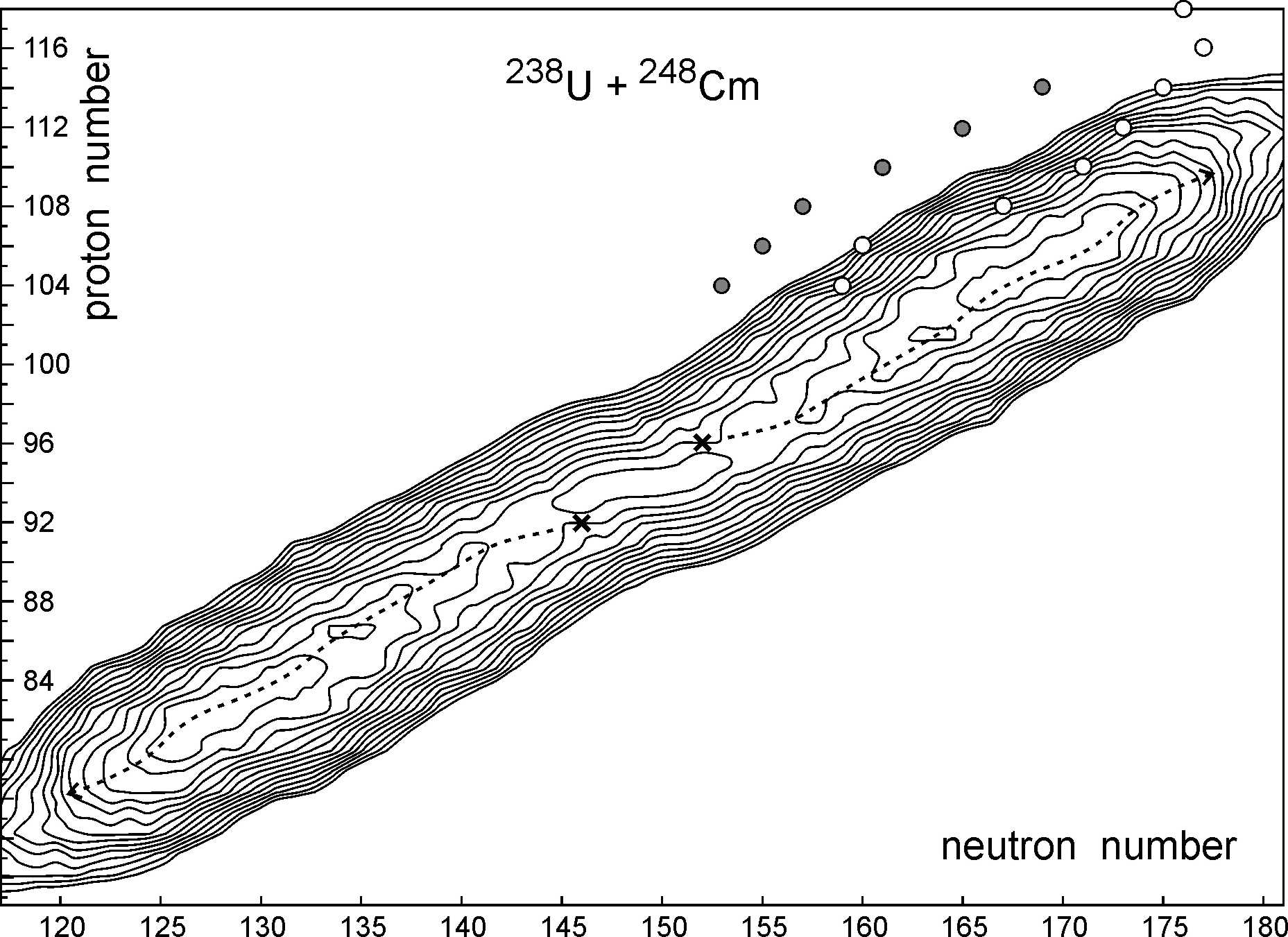} \end{center}
\caption{Landscape of potential energy surface of the nuclear
system formed in collision of $^{238}$U with $^{248}$Cm (contact
configuration, dynamic deformation $\beta_2=0.2$, contour lines
are drawn over 1~MeV energy interval). Open circles correspond to
the most neutron-rich nuclei synthesized in $^{48}$Ca induced
fusion reactions while the filled ones show SH nuclei produced in
the ``cold'' fusion with lead target. The dotted line shows the
most probable evolution in multi-nucleon transfer
process.\label{ucm}}
\end{figure}

We extended our approach taking into consideration neutron and
proton asymmetries separately instead of one mass-asymmetry
parameter used before \cite{ZG07}. The potential energy surface of
the giant nuclear system formed in collision of $^{238}$U and
$^{248}$Cm nuclei is shown in Fig.~\ref{ucm}. It is calculated
within the two-center shell model for a configuration of two
touching nuclei (with fixed value of dynamic deformation
$\beta_2=0.2$) depending on numbers of transferred protons and
neutrons. The initial configuration of $^{238}$U and $^{248}$Cm
touching nuclei is shown by the crosses.

\begin{figure}[h]
\begin{center}
\includegraphics[width = 8.0 cm, bb=0 0 1018 1068]{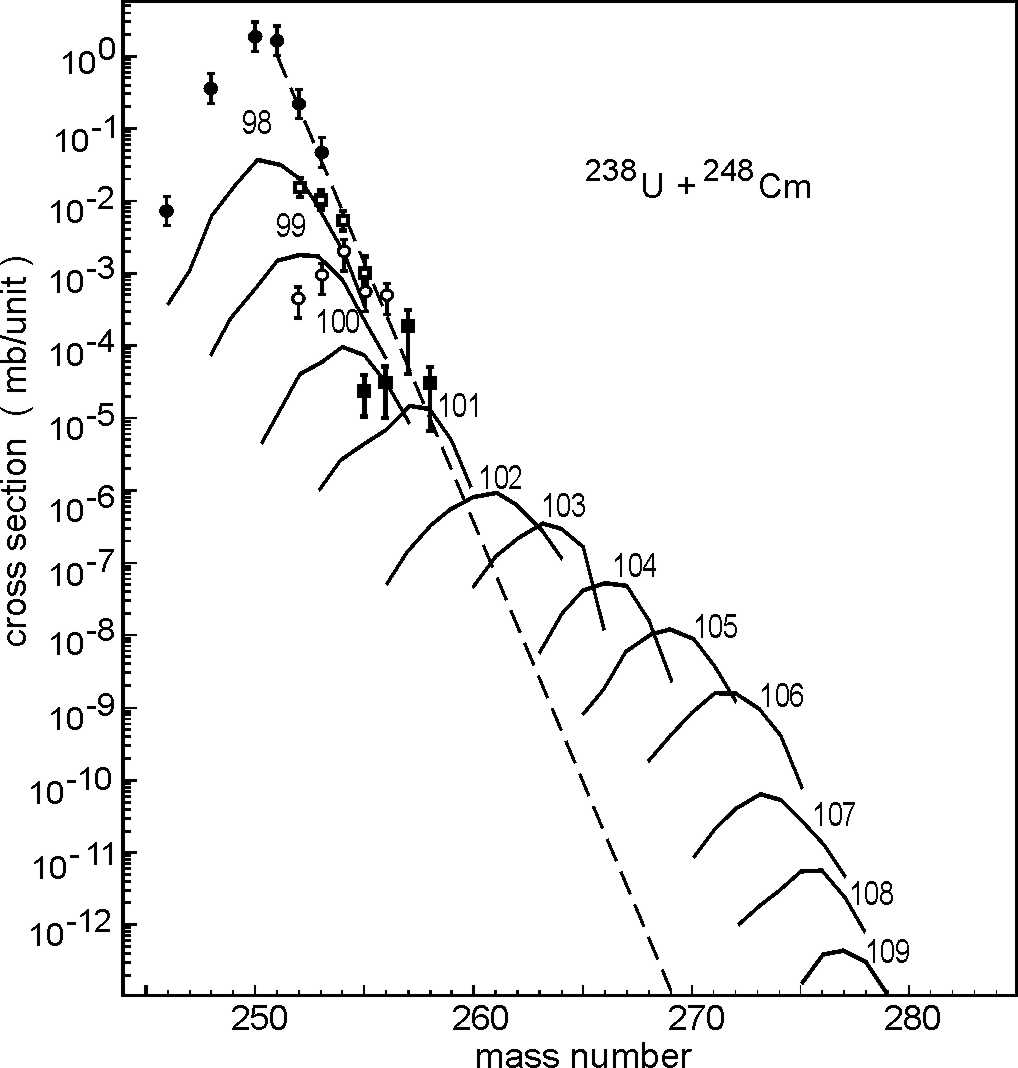} \end{center}
\caption{Yield of survived isotopes of SH nuclei produced in
collisions of $^{238}$U with $^{248}$Cm at 800~MeV center-of-mass
energy. Experimental data for Cf (filled circles), Es (open
rectangles), Fm (open circles) and Md isotopes (filled rectangles)
obtained in \cite{Schadel82} are also shown. Dashed line shows the
expected locus of transfer reaction cross section without the
shell effects.\label{ucm_cs}}
\end{figure}

In low-energy damped collisions of heavy ions just the potential
energy surface regulates to a great extent the evolution of the
nuclear system. From Fig.~\ref{ucm} one sees that in the course of
nucleon exchange the most probable path of the nuclear system
formed by $^{238}$U and $^{248}$Cm lies along the line of
stability with formation of SH nuclei which have many more
neutrons as compared with those produced in the ``cold'' and
``hot'' fusion reactions. Due to fluctuations even more neutron
rich isotopes of SH nuclei may be formed in such transfer
reactions.

The yield of survived SH elements produced in the low-energy
collisions of actinide nuclei is rather low, though the shell
effects give us a definite gain as compared to a monotonous
exponential decrease of the cross sections with increasing number
of transferred nucleons. In Fig.~\ref{ucm_cs} the calculated EvR
cross sections for production of SH nuclei in damped collisions of
$^{238}$U with $^{248}$Cm at 800~MeV center-of-mass energy are
shown along with available experimental data. As can be seen,
really much more neutron rich isotopes of SH nuclei might be
produced in such reactions.

Of course, the reliability of our predictions for the processes
with a transfer of several tens of nucleons is not very high. In
this connection more detailed experiments have to be performed
aimed on the study of the shell effects in the mass transfer
processes in low-energy damped collisions of heavy ions. The
effect of ``inverse'' quasi-fission may be studied also in
experiments with less heavy nuclei. For example, in the collision
of $^{160}$Gd with $^{186}$W we may expect an enhanced yield of
the binary reaction products in the regions of Ba and Pb just due
to the shell effect \cite{ZG07b}. The experimental observation of
this effect and the measurement of the corresponding enhancement
factor in the yield of closed shell nuclei might allow us to make
better predictions (and/or simple extrapolations) for heavier
nuclear combinations which are more difficult for experimental
study.

\section{Conclusion}\label{Conc}

Thus we may conclude that there are several very promising
possibilities for the synthesis of new SH elements and isotopes.
First of all, we may use the titanium beam (instead of $^{48}$Ca)
and actinide targets to move forward up to the element 120. The
estimated EvR cross sections are rather low (at the level of
0.1~pb) but quite reachable at available setups. If the
experiments with titanium beam will confirm our expectations, then
we have to find a possibility to increase the beam intensity and
the detection efficiency (totally by one order of magnitude) and
go on to the chromium and iron beams (aiming to the elements 122
and 124). The use of light and medium mass neutron-rich
radioactive beams may help us to explore and to fill the ``blank
spot'' at the north-east part of the nuclear map. Such a
possibility is also provided by the multi-nucleon transfer
processes in low-energy damped collisions of heavy actinide
nuclei, if the shell effects really play an important role in such
reactions. The production of SH elements in fusion reactions with
accelerated fission fragments looks less encouraging. Only if an
extremely high beam intensity will be attained, the promises are
increasing.

\end{document}